\documentclass{article}

\usepackage{titling}
\title{On Using Micro-Clouds to Deliver the Fog}

\usepackage[sc]{mathpazo} 
\usepackage[T1]{fontenc} 
\usepackage{microtype} 
\usepackage[hmarginratio=1:1,top=25mm,columnsep=20pt]{geometry} 
\usepackage[hang, small,labelfont=bf,up,textfont=it,up]{caption} 
\usepackage{booktabs} 
\usepackage{float} 
\usepackage{paralist} 

\usepackage{abstract} 

\usepackage{titlesec} 
\renewcommand\thesubsection{\Roman{subsection}} 
\titleformat{\section}[block]{\large\scshape\centering}{\thesection.}{1em}{} 
\titleformat{\subsection}[block]{\large}{\thesubsection.}{1em}{} 

\usepackage{fancyhdr} 
\usepackage{amsfonts,amsmath,amssymb,amstext,latexsym}
\usepackage{color}
\usepackage[usenames,dvipsnames,table]{xcolor}
\usepackage{multirow}

\usepackage{graphicx}
\usepackage[caption=false]{subfig}
\newcommand{\fig}[1]{Fig.~\ref{#1}}

\usepackage[pdftex,
            pdfauthor={Yehia Elkhatib},
            pdftitle={On Using Micro-Clouds to Deliver the Fog},
            pdfkeywords={Cloud computing, Fog computing, Micro-clouds, Mini-clouds, Cloudlets, Internet of Things}]{hyperref}

\newcommand*{\sect}[1]{\S\ref{#1}}
\usepackage{xspace}

\newcommand*{\cf}{\textit{cf.}\@\xspace}
\newcommand*{\eg}{\textit{e.g.}\@\xspace}
\newcommand*{\ie}{\textit{i.e.}\@\xspace}
\makeatletter
\newcommand*{\etc}{%
    \@ifnextchar{.}%
        {etc}%
        {etc.\@\xspace}%
}
\newcommand*{\etal}{%
    \@ifnextchar{.}%
        {et al}%
        {et al.\@\xspace}%
}
\makeatother

\begin{document}

\date{}
\title{\thetitle}

\author{
	Yehia Elkhatib,\textsuperscript{$\spadesuit$} 
	Barry Porter,\textsuperscript{$\spadesuit$}
	Heverson B. Ribeiro,\textsuperscript{\textdaggerdbl}\\
	Mohamed Faten Zhani,\textsuperscript{$\star$} 
	Junaid Qadir,\textsuperscript{$\diamondsuit$} 
	and Etienne Rivi{\`e}re\textsuperscript{\textdaggerdbl}\\[4mm]
    \small {$\spadesuit$} Lancaster University, United Kingdom\\
    \small {\textdaggerdbl} University of Neuch{\^a}tel, Switzerland\\
    \small {$\star$} {\'E}cole de Technologie Sup{\'e}rieure, Montreal, Canada\\
    \small {$\diamondsuit$} Information Technology University, Punjab, Pakistan\\
    \normalsize Email: \href{mailto:y.elkhatib@lancaster.ac.uk}{y.elkhatib@lancaster.ac.uk}\\[4mm]
    \textbf{\textcolor{red}{This is a pre-print}}\\
    \textbf{\textcolor{red}{The final version is available on IEEEXplore}}\\
}

\maketitle

\thispagestyle{fancy} 

\begin{abstract}
    Cloud computing has demonstrated itself to be a scalable and cost-efficient solution for many real-world applications. However, its modus operandi is not ideally suited to resource-constrained environments that are characterized by limited network bandwidth and high latencies. With the increasing proliferation and sophistication of edge devices, the idea of fog computing proposes to offload some of the computation to the edge. To this end, \emph{micro-clouds}---which are modular and portable assemblies of small single-board computers---have started to gain attention as infrastructures to support fog computing by offering isolated resource provisioning at the edge in a cost-effective way. We investigate the feasibility and readiness of micro-clouds for delivering the vision of fog computing. Through a number of experiments, we showcase the potential of micro-clouds formed by collections of Raspberry Pi computers to host a range of fog-related applications, particularly for locations where there is limited network bandwidths and long latencies.
\end{abstract}

\section{Introduction}\label{sec:intro}

\begin{figure}[!thb]
	\centering
 	\includegraphics[width=0.6\columnwidth]{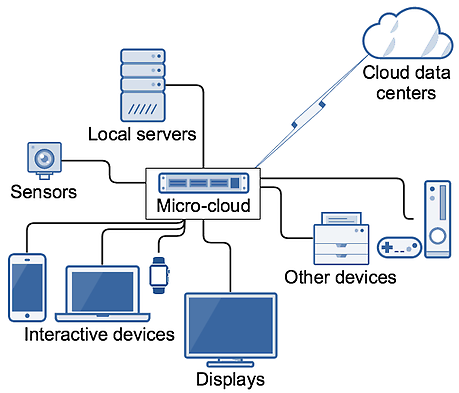}
	\caption{A micro-cloud interconnecting a swarm of edge devices.}
	\label{fig:ex}
\end{figure}

Fog computing is coming. This paradigm allows devices at the edge of the network to become proactive in hosting as well as consuming data and services~\cite{Vaquero2014fog}. 
This has great potential for interconnecting the swarm of such edge devices: wearables, sensors, smart traffic controllers, interactive displays, \etc (see \fig{fig:ex}).
Perhaps more importantly, the fog paradigm offers great potential to provide Internet services in locations that have poor access to network and computational infrastructures, as is the case in many developing regions of the world.

However, little work has been done to date to identify pragmatic means of deploying fog computing solutions. We identify micro-clouds (\sect{sec:what}) as platforms that offer promising opportunities in edge resource provisioning (\sect{sec:why}). We demonstrate through a series of experiments how such platforms are capable enough to support fog solutions (\sect{sec:exp}) and we give an overview of the state of the art (\sect{sec:soa}), charting some short- to medium-term challenges in this regard.

\section{What are Micro-clouds?}
\label{sec:what}

\subsection{Predecessors}
Cyber-foraging and cloudlets have been around for a few years~\cite{satya2009cloudlets,Crowcroft2011}. 
They were first proposed to provide mobile offloading, but through virtualisation technologies that are rather heavyweight for limited-capability, and potentially transient, edge resources. 
For instance, Satyanarayanan \etal proposed in~\cite{satya2009cloudlets} the use of VirtualBox to launch cloudlet images which are accessed through a thin VNC client, and recent works from the same group (\cf~\cite{Ha2013jit}) 
still rely on virtualisation technologies such as KVM and QEMU. On top of the aforementioned disadvantages, using these technologies is not conducive for management and migration of execution units. In addition, virtual machines (VMs) have the tendency to grow into immutable and brittle units that are also costly to transport when migration is necessary. 
We therefore conclude that these cloudlet efforts are quite clearly designed for dedicated (hence static) and powerful servers which might be provided by ISPs. 

Crowcroft \etal proposed \emph{droplets} as an architecture that compromises between the centralisation of the cloud and the opposite extreme decentralisation, termed \textit{the mist}~\cite{Crowcroft2011}. 
Despite explaining high level droplet mechanics and associated trade-offs, the paper did not specify how such droplets would be deployed. We argue that micro-clouds are an ideal deployment vehicle for such a vision, and our experimental results strongly support this.

\subsection{Micro-clouds}
The recent development of small, cheap, low-power computers has prompted a number of new applications, \eg programmable home automation and entertainment systems. 
Several projects and companies have taken advantage of this and started assembling a number of such devices to create small computing clusters for different purposes (\eg racks of Raspberry Pi (rPi) devices such as the PiFace Rack\footnote{\url{http://www.piface.org.uk/products/piface_rack/}}). 
The availability of this hardware, coupled with advancements in virtualisation and hypervisor 
technologies that enable slicing resources between different users to provide isolated environments, have facilitated the advent of \emph{micro-cloud systems}.

Micro-clouds are standalone computational infrastructures of small size that can be easily deployed in different locations. Their scale is in stark contrast to that of the data centres powering the cloud, yet they can offer similar capabilities in the qualitative terms of access to resources in an on-demand, pay-as-you-go fashion.

A clear distinction needs to be made here. 
\emph{Mini-clouds} and \emph{mini-data centres} are terms used by some to refer to modular server racks deployed indoors to provide high computational and storage capabilities. An early example is Sun's data centre in a shipping container\footnote{\url{http://docs.oracle.com/cd/E19115-01/mod.dc.s20/}}. More recent examples are much smaller, around the size of a full- or half-height 19 inch server rack enclosure. 

Micro-clouds also refer to a modular assembling of computers but with the key difference of being easily portable \textit{and} independent of existing infrastructure (\eg a temperature and humidity controlled server room). Consequently, micro-clouds lend themselves to deployment outdoors as well as indoors, and especially in unprepared or hostile environments.

\section{Why Micro-clouds?}
\label{sec:why}

Micro-clouds are becoming increasingly important for their ability to enhance resource provisioning in both \textit{resource poor} and \textit{resource rich} environments. 
We now briefly discuss these two contrasting use cases.

\subsection{Resource Poor Environments}
\label{sec:why:poor}
We posit that modern applications are not only distributed (as Cavage has pointed out~\cite{Cavage2013around}) but also follow a specific mode of operation. 
Over the years, developers have collectively come up with systems that are essentially modern variants of the classical client-server model. Cloud applications predominantly operate in this fashion. Consider the vast majority of web and mobile applications that run on a client device, usually (as specified by the developer) with limited computation responsibilities such as processing or storage. This client communicates with a back end system, typically hosted in a remote data centre, which does the heavy lifting. 

In many parts of the world, including regions where the average income is relatively low, end user devices have become fairly affordable for a large fraction of the population. 
Moreover, such hardware has gradually become increasingly powerful and resourceful. Nevertheless, these clients still heavily rely on the backend (the `server') where it is much easier to scale thanks to the economies of scale leveraged by cloud service providers (CSPs).

\begin{figure}[!htb]
	\centering
 	\includegraphics[width=0.9\columnwidth]{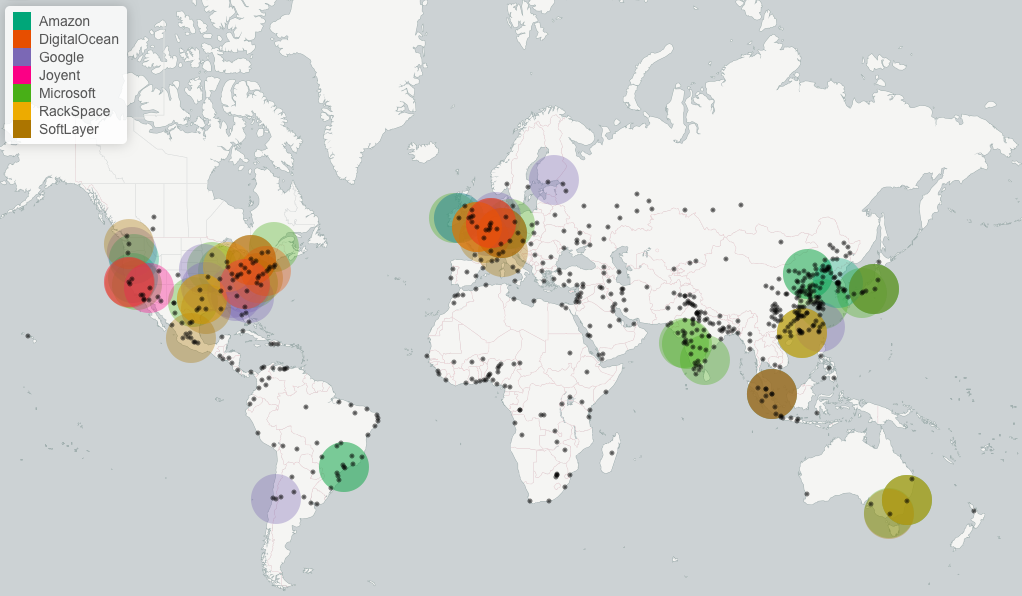}
	\caption{Major cloud data centers (with a 750 mile radius) and major urban areas (black dots).}
	\label{fig:pop}
\end{figure}

Focusing now on the location of the `server', we plot in \fig{fig:pop} the locations of data centers of the major CSPs.\footnote{According to the locations published by the respective CSPs as of Feb 2016 and geocoding through Google Maps API.} 
We also identify the locations of population centers with more than 750,000 people~\cite{nordpil2010pop} with black dots as indicators of significant market potential. 

As is evident from \fig{fig:pop}, for many the `server' is much further away than desired. 
This is amplified by the increasing reliance on interaction between users for business and social purposes. The main concern in this model thus becomes the quality of the network connection between clients and the data centres~\cite{elkhatib2015building}. 

A solution to this problem is to introduce computational resources where needed. 
This is indeed ongoing; \eg Amazon, IBM and Microsoft are planning to build cloud data centers in India~\cite{forbes-aws-india}. 
However, it building data centres in such regions 
is a long term solution involving large budgets, high levels of expertise, and national (even regional) political guarantees. 
In contrast, micro-clouds provide a much lower cost alternative by injecting smaller infrastructures where needed. They require less operational overhead in terms of expertise, energy consumption, and housing, and they do not hinge on large scale geopolitical commitments.

\subsection{Resource Rich Environments}
\label{sec:why:rich}
The Internet of Things (IoT) has brought a lot of promise to various applications such as smart cities, home automation, data-driven industries, \etc
In these environments with a plethora of sensors and actuators, micro-clouds promise a number of opportunities in terms of playing support roles such as aggregation, pre-processing, fault mitigation, and deployment migration assistance. 
This manifestation is supportive of the concept of fog computing~\cite{Bonomi2012fog,Vaquero2014fog}, edge-centric computing~\cite{Lopez2015Edge-centric-Co00}, or the \emph{TerraSwarm}~\cite{lee2014swarm} where heterogeneous devices dispersed around the edge (as opposed to being housed in mega data centres as in the cloud computing model) 
intercommunicate to both provide and consume services. Micro-clouds could be integrated within the fabric of business and residential buildings, and as such bring new computing opportunities via their proximity to, and their (partial) control by, end users.

The easy-to-set-up aspect of micro-clouds renders them highly amenable to emergency situations arising from natural disasters (\eg floods and earthquakes) and security crises (\eg terrorist attacks and riots). 
Resources provided through micro-clouds can be used to set up instant stations to relay safety information, locate victims, coordinate communication between rescue and security units, and provide alternative connectivity means in case long-haul access is lost (\eg mounted on UAVs).

Finally, micro-clouds could be utilised to inject any of the roles of a network middlebox (such as caching and traffic scrubbing) as well as that of software in-network devices (such as an SDN controller or virtualised network function).

\section{Feasibility Experiments}
\label{sec:exp}

We now focus on assessing the feasibility of using the computing capabilities of micro-clouds to deliver fog services. 
We are particularly motivated to ascertain the degree to which such systems could be used to operate isolated execution environments at the edge of the network, thereby supporting the kind of multi-user virtualisation offered in the traditional cloud. 
We zoom in on the rPi device in particular due to its widespread availability and affordability both in terms of purchasing and operation. 
A summary of the different rPi versions used is given in Table~\ref{tab:pis}. All rPi devices used the same Wintec 16GB Class 10 microSD card, with an original adaptor for the older rPi models that only read full size SD cards. 

\begin{table}[hbt]
    \def\arraystretch{1.25}
    \setlength{\tabcolsep}{3pt}
    \centering
    \caption{Raspberry Pi machine specifications.}
    \rowcolors{3}{blue!12}{}
	\begin{tabular}{lccccccc}
 \hline
\multirow{2}{*}{Model} & PCB & \multicolumn{2}{c}{Processor} & \multicolumn{2}{c}{Cache (kB)} & Memory & Power \\
\cline{3-4} \cline{5-6}
                    & Ver. & \#Cores & Clock rate & L1 & L2 & (MB) & (mA) \\
 \hline
    rPi1B & 1.0 & 1 & 700 MHz & 16 & 128 & 256 & 300 \\
    rPi1B & 2.0 & 1 & 700 MHz & 16 & 128 & 512 & 700 \\
    rPi1B+& 1.2 & 1 & 700 MHz & 16 & 128 & 512 & 600 \\
    rPi2B & 1.1 & 4 & 900 MHz & 32 & 512 & 1024 & 800 \\
    rPi3B & 1.2 & 4 & 1.2 GHz & 32 & 512 & 1024 & 800 \\
 \hline
 	\end{tabular}
	\label{tab:pis}
\end{table}

\newcommand{\subfigwidth}{0.325\textwidth}
\newcommand{\subfigwidthwide}{0.43\textwidth}
\begin{figure*}[t!]
\subfloat[50 clients\label{fig:latency:50}]{\includegraphics[width=0.54\textwidth]{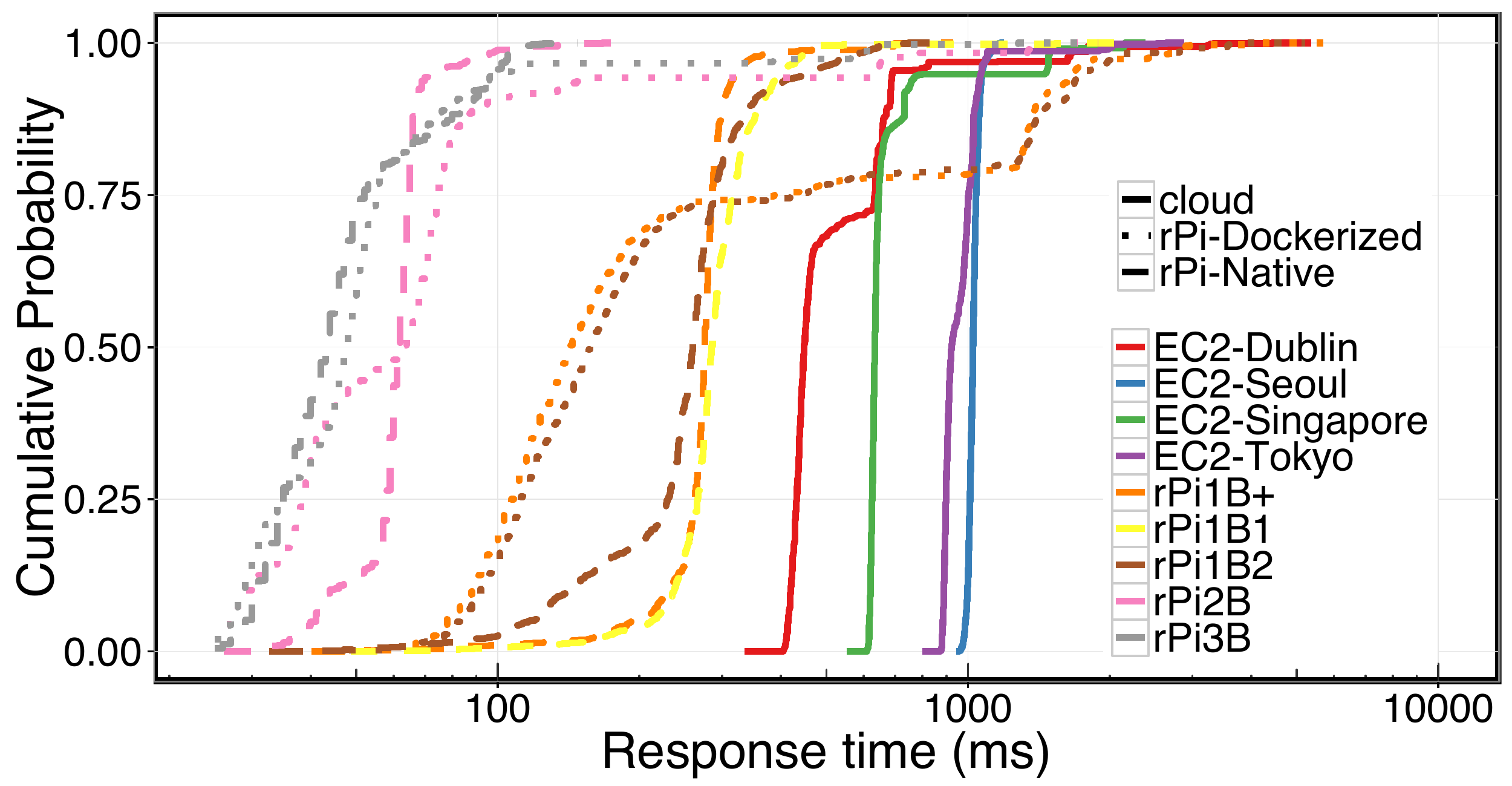}}%
\subfloat[100 clients\label{fig:latency:100}]{\includegraphics[width=0.43\textwidth]{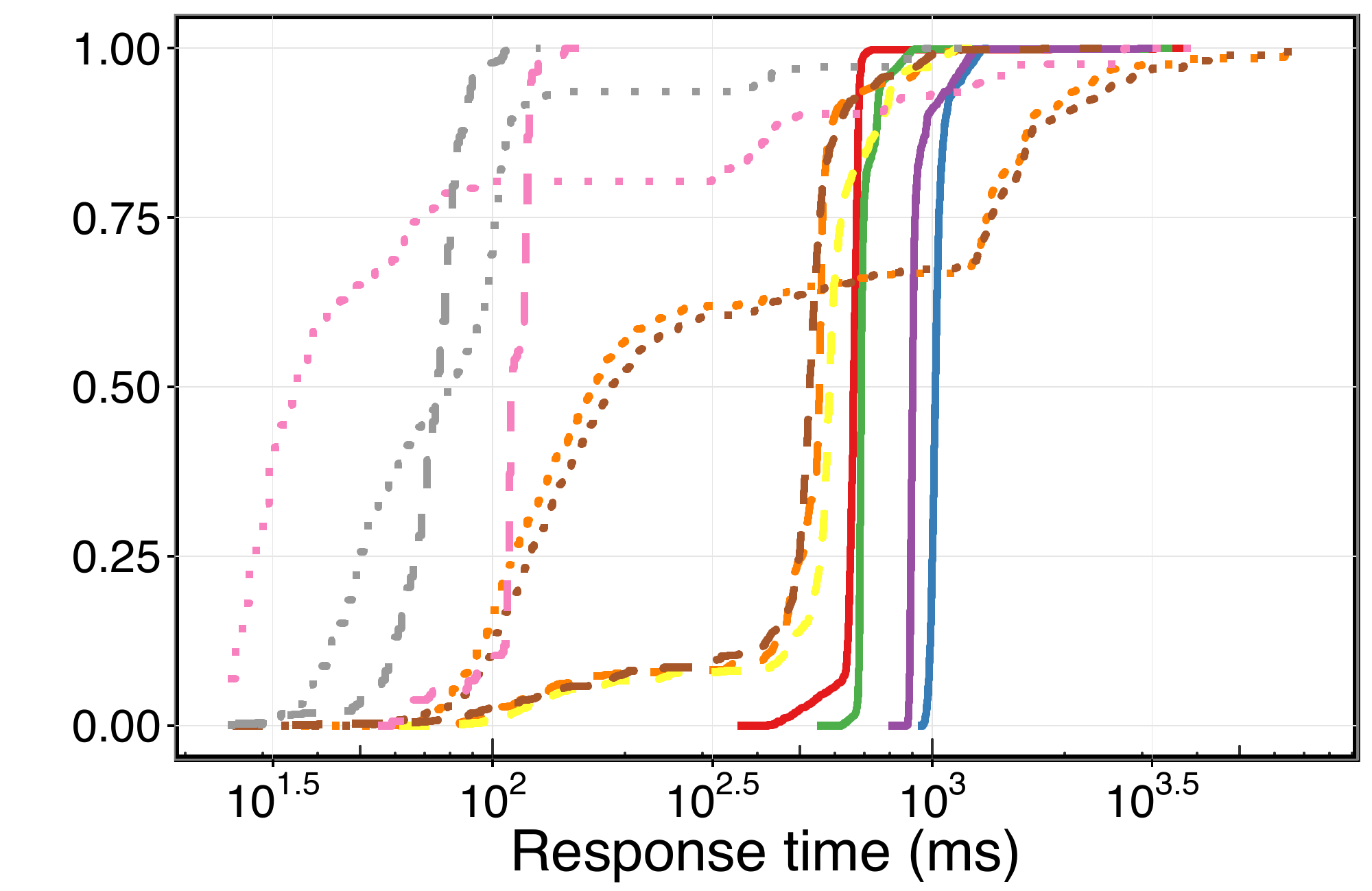}}\\%
\subfloat[150 clients\label{fig:latency:150}]{\includegraphics[width=\subfigwidth]{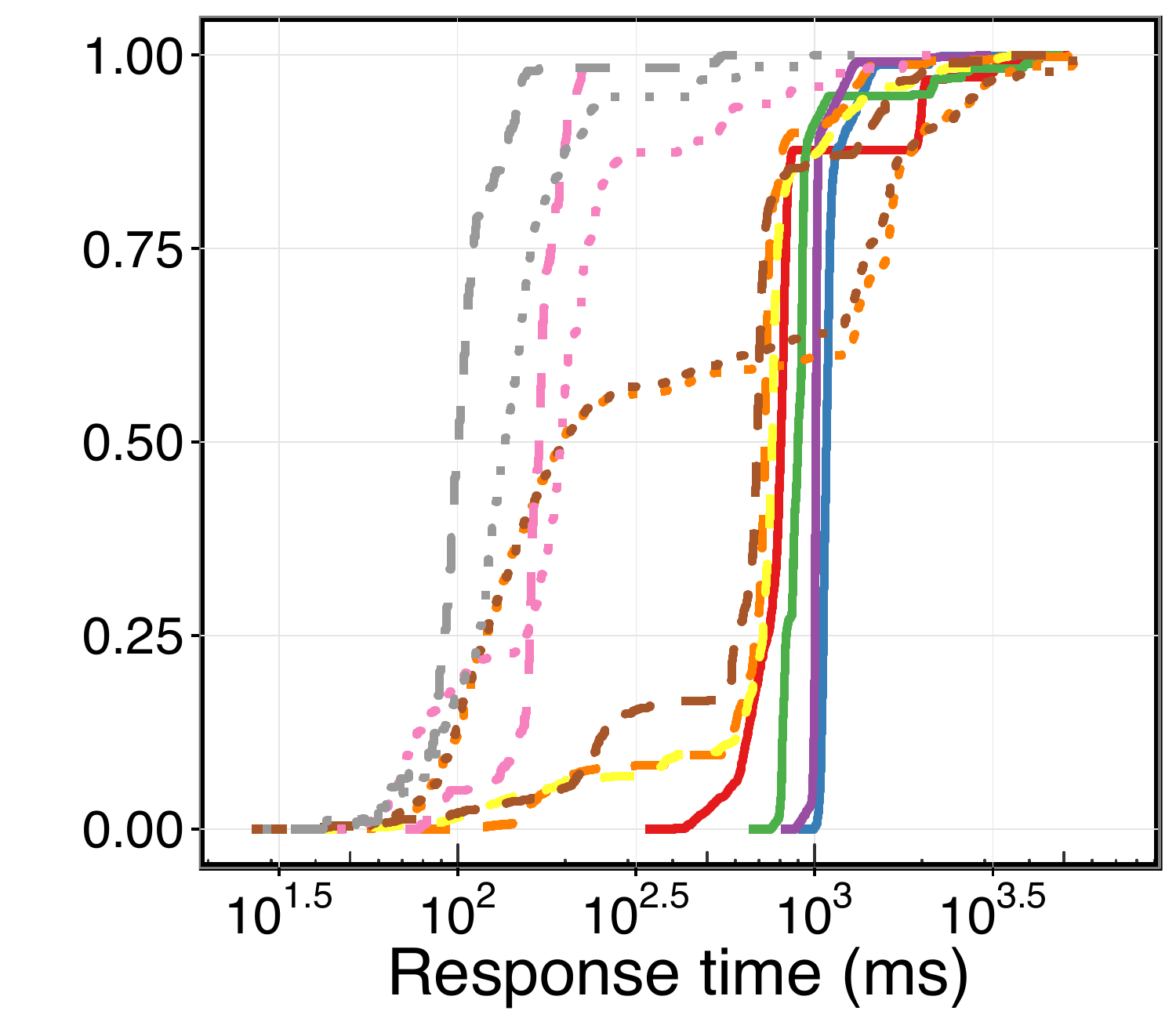}}
\subfloat[200 clients\label{fig:latency:200}]{\includegraphics[width=\subfigwidth]{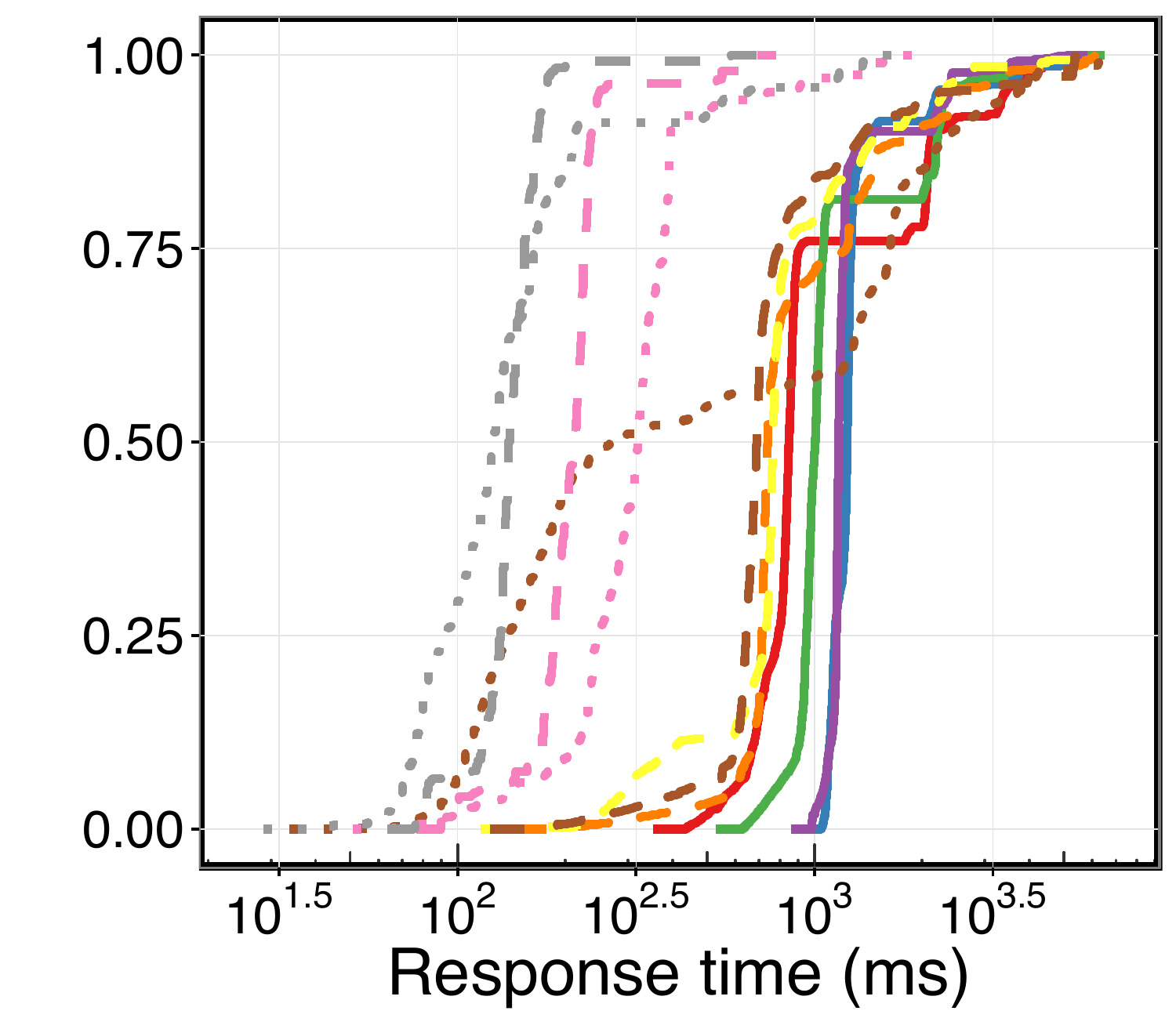}}%
\subfloat[250 clients\label{fig:latency:250}]{\includegraphics[width=\subfigwidth]{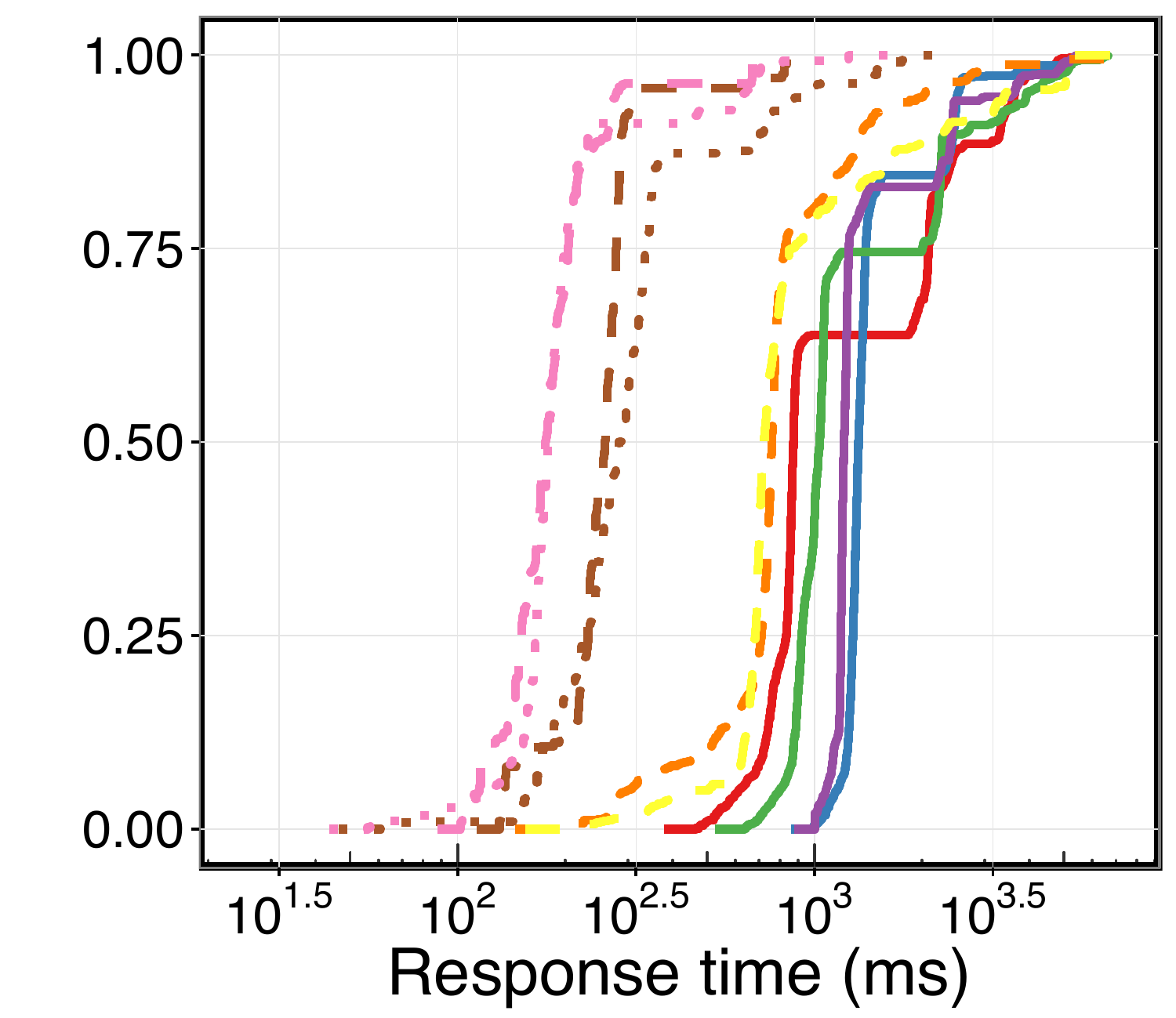}}
	\caption{Page retrieval times for different server setups with varying number of concurrent users.}
	\label{fig:latency}
\end{figure*}

We choose to investigate the ability to run different customer facing servers in Docker containers instead of a classical benchmarking suite. Because the use of hypervisors is unsuitable for lower-power processors without hardware virtualization support, this represents a more realistic lightweight way of achieving isolation in micro-cloud devices. 
The scenario we pick is emblematic of many of the uses planned for edge deployment (\sect{sec:why:poor}) and fog computing (\sect{sec:why:rich}) such as web caching, aggregation, and pre-processing. The use of lightweight containers enables both multi-tenancy and migration readiness which are important in order to cater to changes in user requirements at the edge.

For this we run our rPis using HypriotOS, a Debian-based Linux distribution geared specifically towards running Docker over ARM processors. Unless otherwise stated, we used the Berry release, version 0.7.0 of HypriotOS\footnote{\url{http://blog.hypriot.com/downloads/}}.

In this section, we present results about serving latency (\sect{sec:exp:lat}), hosting capability (\sect{sec:exp:hst}), I/O overhead (\sect{sec:exp:io}), and startup latency (\sect{sec:exp:strt}).

\subsection{Serving Latency}
\label{sec:exp:lat}

Our first experiment investigates the responsiveness of application servers deployed on micro-clouds and their ability to serve a large number of requests. 
We deploy the Apache httpd web server serving a webpage with minimal HTML and one image ($\approx$100kB) in 2 settings: \emph{native}, \ie over Linux, and \emph{Dockerized}, \ie running inside a Docker container. 
We then use the benchmarking tool Apache ab\footnote{\url{http://httpd.apache.org/docs/current/programs/ab.html}} 
to stress test the servers with varying number of concurrent clients between 50 and 250, reaching a total of 1,000 clients per test. 
For these experiments representing an edge deployment, the rPi servers were within 2 hops and $\approx$20ms from the computer simulating clients using ab. 
The results are plotted in \fig{fig:latency}. 

We also deployed a similar server setup (only native httpd) on Amazon EC2 in several of their global data center locations and ran the ab tests from Lahore, Pakistan. These cloud servers were located in the following locations offered by Amazon Web Services: Dublin, Ireland; Seoul, South Korea; Singapore; and Tokyo, Japan. The results are also presented in \fig{fig:latency} (keys starting with ``EC2-'').

We draw the main observations from this set of results.
First, the results confirm that \textit{a classic cloud deployment is not ideal for all scenarios}, especially for end users in locations like Pakistan where data centres (even Asia-Pacific ones) are at a considerable network distance. In such situations, micro-clouds provide a suitable substitute for certain applications requiring low latency. 
Second, most rPis are very capable of handling a significant number of web requests. Despite their limited computational capability, using them in such locations \textit{improves latency by at least an order of magnitude}. However, as the number of concurrent users becomes significantly high, a hybrid deployment solution leveraging both remote cloud data centers and micro-clouds becomes a more viable option as the micro-cloud computational limitations start to show.
Third, running the servers within Docker introduces a notable amount of overhead attributed to isolation. In the case of the Pi1B1, the oldest of the rPis, the native server was able to serve 250 concurrent users while the Dockerized server was unable to serve more than 30 concurrent users.
As a final comment, we note that successive rPi generations are becoming increasingly able to withstand additional load. 

\textbf{Take away:} \textit{A modern rPi is an adequate hosting environment for edge web servers, especially as a replacement for a cloud-hosted VM in a remote data center.}

\subsection{Hosting Capability}
\label{sec:exp:hst}
In our second experiment we explore hosting expanse, \ie we identify the limits of hosting multiple Docker containers on an rPi. 
A presentation at Dockercon 2015\footnote{\url{https://blog.docker.com/2015/10/raspberry-pi-dockercon}} demonstrated that a rPi can run hundreds of httpd servers simultaneously. Although our experiments verified this, we found that containers become practically unusable at these levels of load because the Docker daemon eventually starts pushing newly created containers to virtual memory. While it is therefore technically possible to run as many containers as your virtual address space can accommodate, in reality containers will begin to require an insignificant delay before becoming responsive as the amount of memory paging on and off the SD card increases. Additionally, the rPi overall becomes rather unresponsive at this level of loading. Instead of examining the technical maximum container count, we instead therefore set out to find out the real limit beyond which additional Docker containers become excessive.

We also examine how different services behave in these terms, rather than looking at a single example service. If micro-clouds are to become used as an everyday cloud analogue they will be required to handle a wide variety of services including load balancers, web servers, caches, messages queues, databases, and so on. Our evaluation therefore includes a look into how different services would operate on micro-clouds, covering the canonical applications listed in Table~\ref{tab:apps}. All selected applications were packaged for running on HypriotOS.

\begin{table}[hbt!]
    \def\arraystretch{1.25}
    \setlength{\tabcolsep}{3pt}
    \centering
    \caption{Hosted applications and their types.}
	\begin{tabular}{ll}
 \hline
 Application name & Application Type \\
 \hline
    CrateDB & Database\\
    httpd & Web server\\
    Jenkins & Automation \& integration\\
    Ruby & Programming runtime\\
    ZooKeeper & Configuration \& synchronization\\
 \hline
 	\end{tabular}
	\label{tab:apps}
\end{table}

Our experiments here incrementally deploy as many Docker containers as possible, one at a time, up to a maximum memory threshold that can be used by all deployed containers. The rationale behind this threshold is to preserve a certain amount of memory to the applications themselves and to the operating system. We then monitored the memory utilization of the applications, ending each experiment when this threshold is reached. We set the threshold to 50MB in our experiments.

\begin{figure}[t]
	\centering	
      \includegraphics[width=0.675\columnwidth]{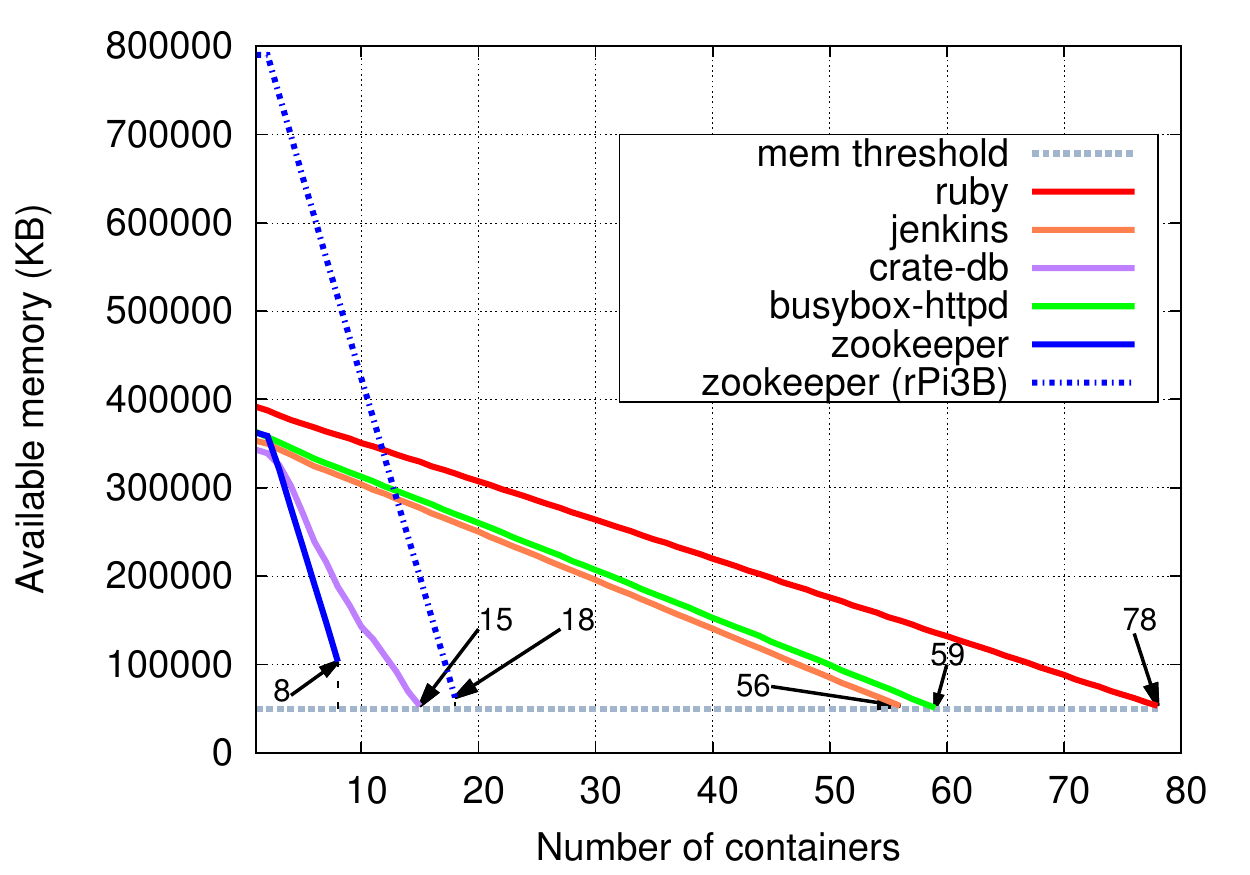}
	\caption{Mean memory usage as a function of the number of containers deployed (10 runs).}
	\label{fig:hosting}
\end{figure}

Contrary to expectation, the rPi's secondary storage capacity (the SD card) did not have a significant impact. For instance, deploying about 60 containers of the httpd server requires only 8.3KB of disk storage. However, the experiments revealed that main memory is a significant bottleneck (\fig{fig:hosting}). 

These experiments were performed on rPi1B2 and rPi1B+ where the maximum memory available to user space applications is around 430MB. The memory available at the start of each deployment was around 380MB. As applications have different memory requirements, the maximum number of containers varies across deployments: from 8 in the case of ZooKeeper to 78 for Ruby. 
We also note that ZooKeeper deployments never reached the threshold, stopping well before this. The explanation for that is straightforward, the amount of available memory before reaching the threshold is not enough to deploy another single container. 
As rPi1B2 and rPi1B+ have the same amount of memory the results in \fig{fig:hosting} depict the performance for both models. 

We repeated the same experiments using other rPi models with improved hosting capability (\ie rPi3B\footnote{For this we needed to switch to the Barbossa release, version 0.8.0.}). In this model the maximum memory available to user space application was around 800MB. The dashed blue line shows that a rPi3B can deploy more than twice as many ZooKeeper containers than rPi1B2 and rPi1B+ models. 

\textbf{Take away:} \textit{An rPi is capable of hosting a significant number of containers at the same time, while preserving their responsiveness.}

\subsection{I/O Overhead}
\label{sec:exp:io}

Our third experiment dives a little deeper into the performance that the rPi architecture delivers to the applications that it hosts. Besides CPU speed, one of the most significant differences in hardware architecture is the physical memory design. Memory access also represents one of the major costs involved in a range of Internet services, from databases to serving resources and processing over small volumes of in-memory data, and so is useful to measure in isolation. We therefore examine the relative cost of reading from and writing to different kinds of memory with our various rPi models, compared to a cloud server. To do this we wrote a program which reads and writes increasingly large amounts of data to secondary storage, and also writes increasingly large amounts of data in RAM (we consider main memory reads and writes to be symmetrical). 

\renewcommand{\subfigwidth}{0.33\columnwidth}
\begin{figure}[!ht]
	\centering	
\subfloat[Write to disk\label{fig:io:disk-write}]{\includegraphics[width=\subfigwidth]{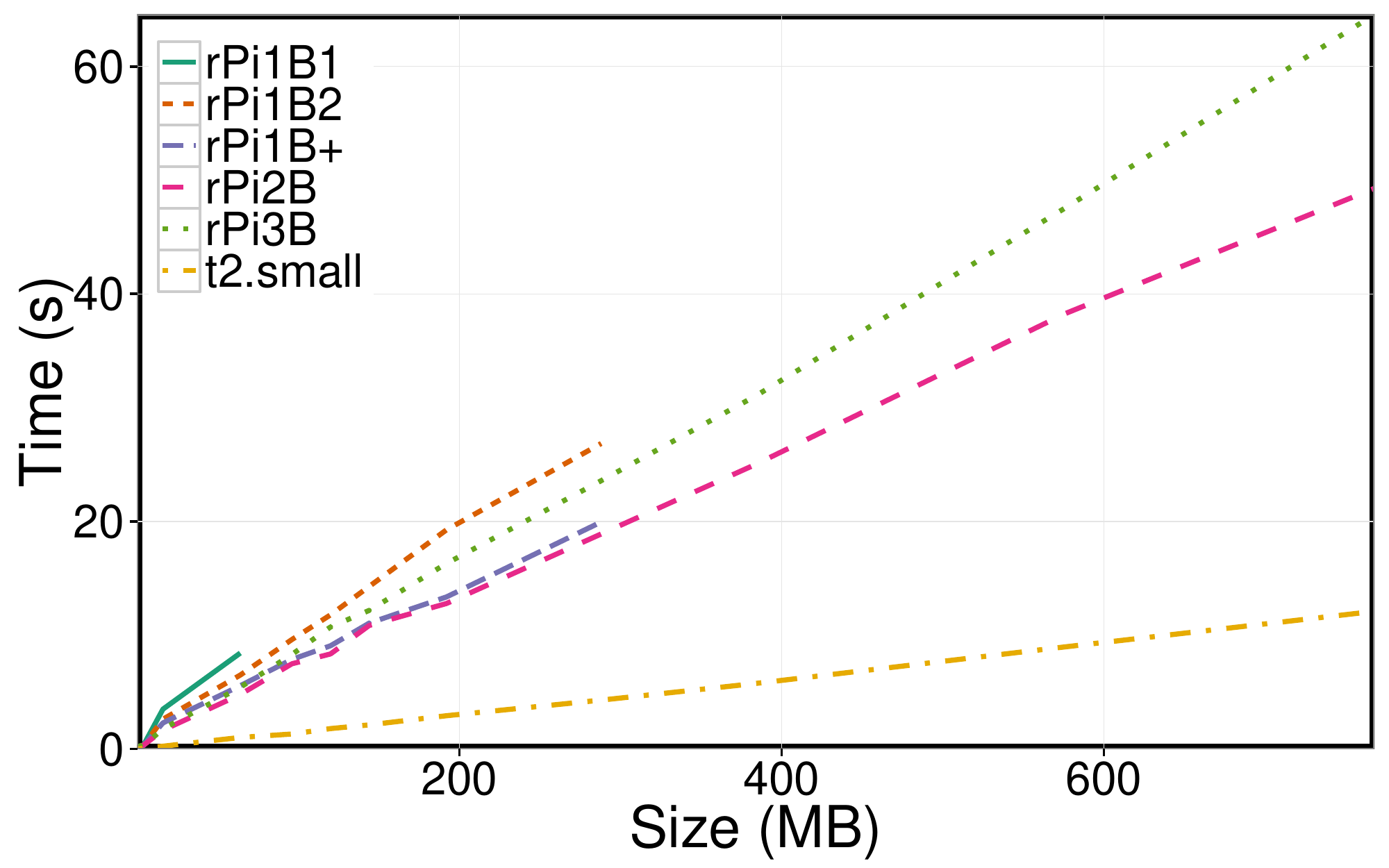}}%
\subfloat[Read from disk\label{fig:io:disk-read}]{\includegraphics[width=\subfigwidth]{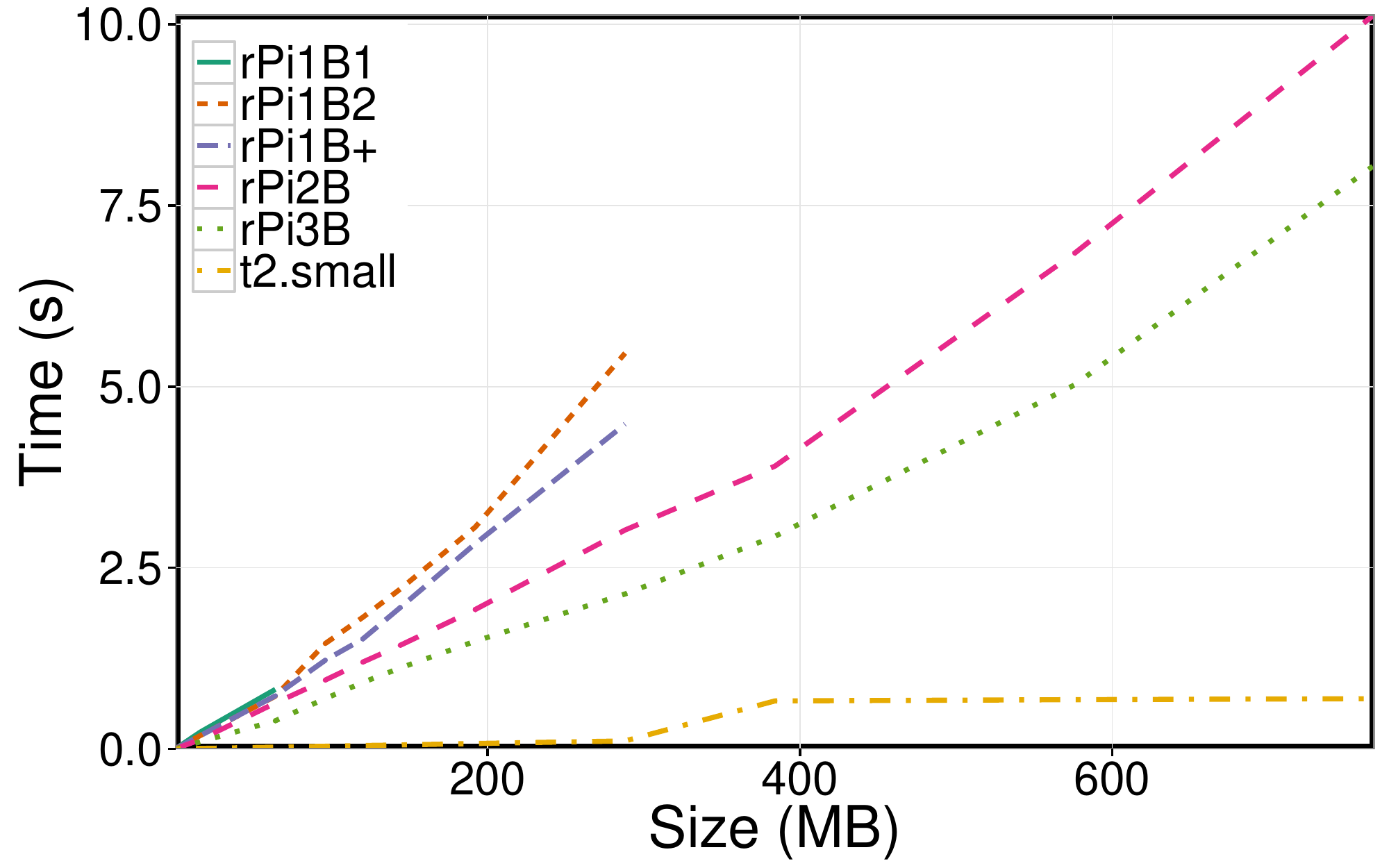}}%
\subfloat[Access to memory\label{fig:io:memory-access}]{\includegraphics[width=\subfigwidth]{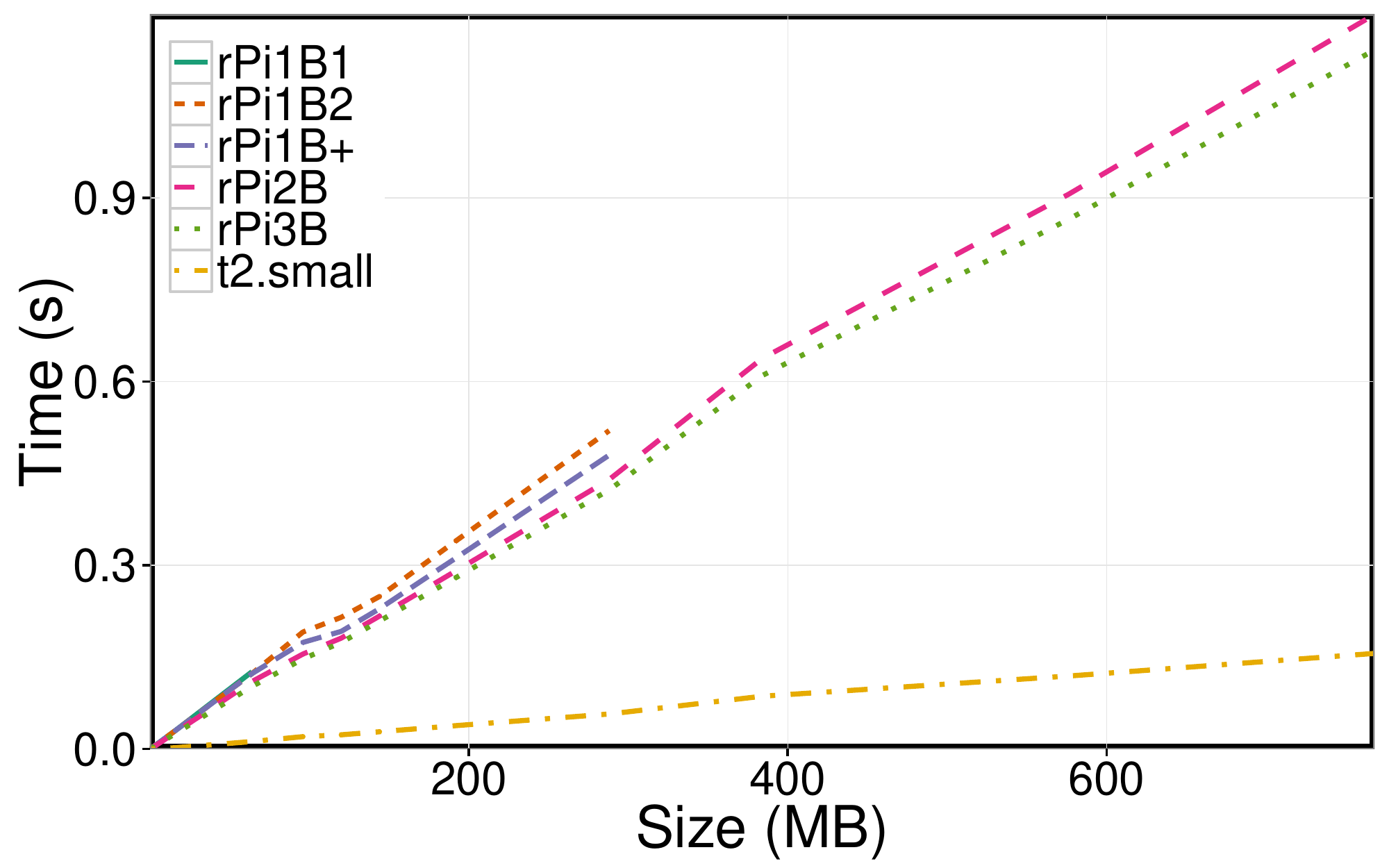}}%
	\caption{Mean disk and memory access times (20 runs).}
	\label{fig:io}
\end{figure}

The results, depicted in \fig{fig:io}, demonstrate the following. 
First, cloud server-based I/O speeds are considerably faster across all types of memory access; this is to be expected due to the generally higher specification of server hardware (\ie CPU speed and cache architectures, system bus speed and main memory latency). However, successive rPi models provide incremental improvements in I/O speeds. 
Second, writing to disk contains a far higher relative penalty across all rPi models than in the server case (compared to disk read or memory access speeds).
We assume that the relative additional disk write latency in rPi systems is caused by the relatively slow write speeds on flash memory used in SD cards~\cite{Chen2009flash}. Software deployments that predominantly use disk reads and memory access, avoiding disk writes, may therefore be even more valuable for efficiency on these devices than on cloud-based servers -- otherwise the network latency gains reported in \sect{sec:exp:lat} may be eclipsed by slower disk access. Of the applications tested in \sect{sec:exp:hst}, the httpd, Ruby and ZooKeeper types are therefore most likely to be suited to the rPi environment. Emerging trends in system design such as entire databases stored in main memory may also be particularly useful, though obviously main memory is capacity-limited. In the wider research picture, because disk access of applications may not be predictable ahead of time, work on adaptive systems may further help to gain the best balance between traditional- and micro-cloud deployments, selecting the optimal placement of a server based on real-time observations.

\textbf{Take away:} \textit{Writing to disk is extremely expensive, but progressively getting better for newer rPi models.}

\subsection{Startup Latency}
\label{sec:exp:strt}
\begin{figure}[thb]
	\centering	
    \includegraphics[width=0.65\columnwidth]{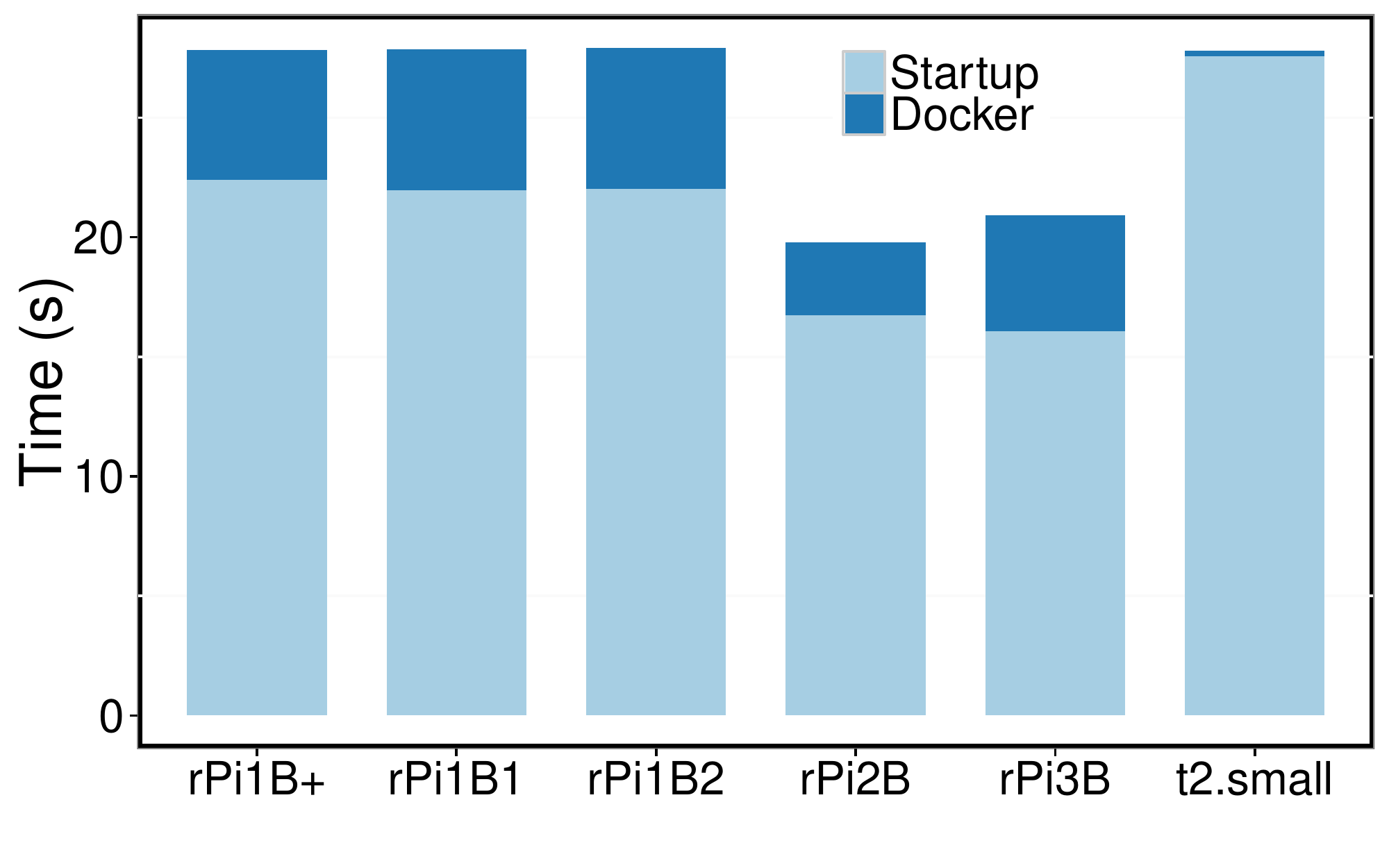}
	\caption{Startup times for different server setups (mean values from 20 tests).}
	\label{fig:startup}
\end{figure}

Our final experiment is to measure the booting time of the rPi devices. This delay is of importance for deployments where electricity shortage is a chronic problem. We present in \fig{fig:startup} the time taken by each of our rPis (running HypriotOS) 
to be ready to interact with and also to start a pre-downloaded httpd Docker container. 
We also include the time it takes to start a t2.small EC2 instance running Ubuntu 14.04 (\emph{without} network delay) as a baseline.

All rPi models take less time to start than an EC2 VM. In fact, the more recent models (rPi2B and rPi3B) undercut the EC2 VM by over 10 seconds, a 40\% margin. Starting Docker on the rPis takes considerably more time, though. However, even with this additional delay, one would have a running Docker container on a rPi2B or a rPi3B up to 8 seconds before an EC2 VM is made ready. Note that these are pure OS and hypervisor latencies without accounting for network latency discussed in \sect{sec:exp:lat}, which would tip the advantage in rPi's favour even further.

\textbf{Take away:} \textit{rPis boot quickly, making them suitable for dynamic deployments even where power is sporadic.}

\section{State of the Art}
\label{sec:soa}
Having empirically explored their feasibility, we now consider the state of the art in micro-cloud implementations and promising future directions. Work in the micro-cloud domain depends on the interplay between 3 main axes: hardware, resource management, and programming abstractions. We survey each area in turn. 

\subsection{Hardware}
The small single-board computers that are the building blocks of micro-clouds are advancing all the time with better chips, additional modules, lower power requirements, and smaller size. Aside from such developments, the major challenge here is to assemble such devices to build the micro-clouds. Several commercial and research ventures have already started to work on this using different strategies. We highlight a few here.

Micro-clouds can be deployed in a number of business and mission-critical application in hazardous or remote locations for applications such as military tactics~\cite{warr2014mobile} and disaster recovery~\cite{Lenk2014CloudStandby}. Several micro-cloud deployments and hardware exist, of which we highlight a few.

\begin{itemize}

    \item \textit{Viacont Automotive Linux Box}\footnote{\url{http://www.viacont.com/}} is a small ruggedised enclosure containing an rPi with different connectivity options (WiFi, 3G, Bluetooth) and on-board sensors (thermometer, accelerometer). 
    
    \item \textit{RuggedPOD}\footnote{\url{http://ruggedpod.qyshare.com/}} is an Open Hardware project to provide ruggedised, energy efficient enclosures that can house up to 4 microATX motherboards immersed in dielectric oil for cooling and controlled thermal insulation.
    
    \item The \textit{Micro Weather station}\footnote{\url{http://www.inveneo.org/2015/09/how-to-design-a-data-center-for-the-developing-world/
}} is a solar powered, outdoor-ready enclosure which includes a Banana Pi and a battery.

    \item \textit{Idein's Actbulb}\footnote{\url{http://actbulb.idein.jp/}} is an rPi-based device that plugs in light sockets, equipped with a camera and a microphone to collect and analyse audiovisual data to be used by different applications via WiFi. 
    The same Japanese company is also working on bundling together 16 Pi Zero chips on one board, a product called PiZero Cluster\footnote{\url{https://twitter.com/9_ties/status/689707306494271488}}.

    \item \textit{Iridis-Pi} \cite{Cox2013iridis} assembled 64 rPis in a custom enclosure made from Lego blocks. Iridis-Pi is advertised as an alternative to virtualisation 
for teaching distributed computing concepts without access to high-end HPC clusters. Iridis-Pis run the Raspbian operating system. 
Costing around \$4,800 including gigabit switches, the Iridis-Pi cluster provides relatively cost-effective computational power was not designed for portability: each Pi is individually wired using Ethernet and microUSB power cables, resulting in cluttered and fragile interconnection. The cluster also relies on ambient cooling, so it cannot adapt to variations in its environment. A similar effort emerged from the legacy of the Beowulf cluster~\cite{kiepert2013beowulf}. This contained 32 rPis running Arch Linux, costing a total of \$1,967 including 4 enclosure cooling fans. This cluster suffers from the same wiring problem as Iridis-Pi.

    \item The \textit{Parallella} cluster\footnote{\url{http://www.parallac.org/}} is made up of 8 credit-card sized boards with 18-cores housed with a gigabit Ethernet switch inside a PVC tower (6'' in diameter, 13'' high), cooled by a fan mounted at the top. The base contains 2 Intel NUCs and their connections which provide 32GB of RAM, 240GB of storage, and WiFi connectivity. Parallac-X1, a modified version, featured a similar architecture but with 16 boards in a cubical enclosure. 
    BitScope\footnote{\url{http://www.bitscope.com/product/blade/?p=about}} and PicoCluster\footnote{\url{http://www.picocluster.com/}} offer alternative commercial assemblies.

\end{itemize}

\subsection{Resource Management}
Typical infrastructure management tools (\eg OpenStack) have been developed for the cloud, but are far too heavyweight for micro-cloud infrastructures. Recent technologies such as containers (\eg Docker\footnote{\url{https://docker.com/}}, rkt\footnote{\url{https://coreos.com/rkt}}, ContainerX\footnote{\url{http://containerx.io/}}, and MultiBox~\cite{hadley2015multibox}) and Unikernels~\cite{madhavapeddy2013unikernels} provide lightweight alternatives that are more appropriate for (re)deploying applications as they are much more efficient in sharing resources at the OS level compared to hardware resource virtualisation by hypervisors. They are also flexible and easily controllable through offering good exposure to the developer on the outside.

Consequently, ongoing efforts to design operating systems and orchestration tools suitable for this scale of computers are revolving around container technologies and the microservices architectural philosophy. HypriotOS is a leading effort here, yet it is still rather bloated and general purpose; we encountered some instabilities whilst carrying out our experiments albeit infrequent. 
There is room for a leaner OS, and for others dedicated for running not just Docker containers but also complimentary tools such as Swarm (for clustering hosts) and Compose (to orchestrate multi-container deployments). 
Operating minimal isolated application stacks using technologies other than Docker, \eg Unikernels, would also open up room for further development.

\subsection{Programming Models}
A few works have been undertaken to build common tools and vocabulary to simplify the setting up and operation of fog systems. This is especially important for developing for and deploying across different fog devices, and to link them to the cloud~\cite{Elkhatib2016crosscloudmap}.

Mobile Fog~\cite{Hong2013mobilefog} defines a specification for location-aware applications. The model assumes edge deployment environments where on-demand resources are manageable through IaaS-like APIs. It prescribes a set of event handlers that an application must implement to be able to exploit said resources in order to scale and relocate in response to user requirements. 
Zhang \etal \cite{Zhang2015enough} introduce a data-centric abstraction API based around distributed logs accessible through names not locations. 
The Holon architecture~\cite{Blair2015holons} is a more generic, conceptual framework to support the composition of systems-of-systems. Holons combine a declaratively defined structure with a service implementation that supports opportunistic post-deployment composition with other holons based on functional and non-functional requirements.

Deri and Fusco~\cite{Deri2013ntm} present an architecture for the management of network monitoring data streams. It describes a deployment of micro-clouds to diffuse network flow analysis and perform some of it towards the edge rather than all going back to the core of the network. Micro-cloud resources are used to reduce delay between analysing network flows (for intrusion and anomaly detection) and responding accordingly (by implementing policies). 
Others (\eg CHive~\cite{Theeten2015chive}) use micro-clouds to pre-process and filter large datasets, such as realtime sensor data, before communicating the results to the cloud. 
More solutions are required in this direction to support designing applications that can be readily divisible between multiple deployment infrastructures. For instance, an application's presentation layer might reside on several multi-cloud instances close to the users, but might share a data tier managed in a cloud data center and other micro-clouds. 

As for the commercial sector, most contributions have focused on integration frameworks that ease the setup of machine-to-machine communications in between colocated devices and with cloud server. In turn, this simplifies the development and maintenance of solutions such as home automation, personal healthcare, and smart traffic systems. Examples here include Arkessa, Axeda, Lyric, Resin, SmartBear, Thingsquare, and Withings.

\section{Concluding Remarks}
\label{sec:disc}

The fog promises to bring the cloud nearer to the edge and enable edge devices to intercommunicate in a richer manner, which is of importance to both resource rich and resource poor environments. However, this builds a need for additional resource provisioning between the cloud and the ever growing end user applications at the edge. In micro-clouds we have found considerably capable and affordable edge infrastructures that are extremely cost-effective considering their low cost, small retail estate, and low power requirements.

The results have shown us that using Docker on rPi enables us to sustain a significant amount of isolated services (web servers, database servers, and application runtimes, \etc) that are able to serve a large number of requests. This benefits both applications that are resource-rich (such as a smart city IoT deployment) and those that are resource-poor (such as a community in a remote location). Our network latency results demonstrate significant potential for moving services closer to the consumer using cheap, micro-service deployments, as long as the volume of co-hosted services and the type of service (avoiding those with heavy disk access) are carefully considered.

We have experimented with a range of different generations of the rPi computer boards. We are not certain whether our findings would be generalisable to other small computers (\eg Pine64, LattePanda) that can be used for assembling micro-clouds. This is a question we are keen to answer in the near future. 
What is more pertinent at this stage, however, is how such extremely cost-effective utility from one device would be substantially amplified when considering a micro-cloud system composed of a host of devices such as the rPis we tested.

 \section*{Acknowledgments}

This work has been partially funded by CHIST-ERA under the DIONASYS project: the UK's Engineering and Physical Sciences Research Council (EPSRC) grant reference EP/M015734/1, and Swiss National Science Foundation (SNSF) grant reference 155249; and also by the UK's EPSRC under grant reference EP/M029603/1.

\bibliographystyle{ieeetr}
\bibliography{microclouds}

\end{document}